\documentclass[aps,prb,showpacs,superscriptaddress,twocolumn]{revtex4-1}
\usepackage{latexsym}
\usepackage{graphicx,epstopdf}
\usepackage{amsfonts}
\usepackage{amssymb}
\usepackage{amsmath}
\usepackage{bm}
\usepackage{braket}
\usepackage{color}
\definecolor{lightblue}{rgb}{0.13, 0.26, 0.99}
\usepackage[
colorlinks=true,
urlcolor=lightblue,
citecolor=lightblue,
linkcolor=lightblue,
hyperfootnotes=false]{hyperref}
\newcommand{\Tr}{\operatorname{Tr}}

\begin{document}
\title{
Dimerizations in spin-$S$ antiferromagnetic
chains with three-spin interaction
}
\author{Zheng-Yuan Wang}
\affiliation{Department of Physics, Tokyo Institute of Technology,
Oh-okayama, Meguro-ku, Tokyo, 152-8551, Japan}
\author{Shunsuke C. Furuya}
\affiliation{DPMC-MaNEP, University of Geneva, 24 Quai Ernest-Ansermet
CH-1211 Geneva, Switzerland}
\author{Masaaki Nakamura}
\affiliation{Max-Planck-Institut f\"{u}r Festk\"{o}rperforschung,
Heisenbergstrasse 1, D-70569 Stuttgart, Germany}
\affiliation{Institute of Industrial Science, the University of Tokyo,
Meguro-ku, Tokyo, 153-8505, Japan}
\author{Ryo Komakura}
\affiliation{Department of Physics, Tokyo Institute of Technology,
Oh-okayama, Meguro-ku, Tokyo, 152-8551, Japan}
\date{published 23 December 2013}
\begin{abstract}
We discuss spin-$S$ antiferromagnetic Heisenberg chains with three-spin
interactions, next-nearest-neighbor interactions, and bond alternation.
First, we prove rigorously that there exist parameter regions of the exact dimerized
ground state in this system.
This is a generalization of the Majumdar-Ghosh model to arbitrary $S$.
Next, we discuss the ground-state phase diagram of the models
by introducing several effective field theories and the universality classes of
the transitions are described by the level-$2S$ $\mathrm{SU}(2)$
Wess-Zumino-Witten model and the Gaussian model.
Finally, we determine the phase diagrams of $S=1$ and $S=3/2$ systems by using exact
diagonalization and level spectroscopy.
\end{abstract}
\maketitle
\section{Introduction}
\label{sec.intro}

Models with exact ground states have provided the foundations
for investigating strongly correlated systems in a
nonperturbative manner. The Affleck-Kennedy-Lieb-Tasaki
(AKLT) model with an exact valence-bond-solid (VBS)
ground state is a good example of such systems.~\cite{AKLT}
It is now widely accepted that the VBS state captures the qualitative
nature of the Haldane phase in the integer-$S$
antiferromagnetic Heisenberg (AFH) chain.
For $S=1/2$ systems, on the other hand,
the Majumdar-Ghosh (MG) model is known to have a fully dimerized state
as the exact ground state.~\cite{MG}
It has been shown that this state captures the dimer phases that appear
in certain parameter regions of the frustrated AFH chain,~\cite{Okamoto-N}
and experimentally observed states such as in CuGeO$_3$.~\cite{Hase-T-U}
An extension of the exact dimerized ground state to higher
dimensions has also been investigated, because it is relevant
to experimental studies, such as those of the two-dimensional
Shastry-Sutherland model~\cite{Shastry-S}
compound SrCu$_2$(BO$_3$)$_2$.~\cite{Kageyama}

In this paper, we consider an intriguing model which has
a dimerized exact ground state in certain parameter regions:
a general-$S$ spin chain including three-spin, next-nearestneighbor
interactions, and bond alternation,
\begin{align}
 \mathcal H
 &=
 J_1 \sum_i \{1 - \delta (-1)^i \} \bm S_i \cdot \bm S_{i+1}\nonumber\\
 &+J_2\sum_i \bm{S}_{i-1} \cdot \bm{S}_{i+1}\nonumber\\
 &+J_3\sum_i 
 \bigl[ (\bm S_{i-1} \cdot  \bm S_i)(\bm S_i \cdot \bm S_{i+1})
 +\mathrm{H.c.}\bigr],
 \label{eq:dJ1J2J3}
\end{align}
where $\mathrm{H.c.}$ denotes the Hermitian conjugate.
The model \eqref{eq:dJ1J2J3} is schematically described in Fig.~\ref{fig:dJ1J2J3}.
As we will see below, our model \eqref{eq:dJ1J2J3} may be regarded as a natural,
but not naive,
extension of the  $S=1/2$ MG model to a general-$S$ case.
Note that, just as for several extensions of the MG model,~\cite{Okamoto-N,Takano,Nakano-T}
our model \eqref{eq:dJ1J2J3} has longer-range interaction than nearest neighbor.
\begin{figure}[b!]
 \centering
 \includegraphics[width=\linewidth]{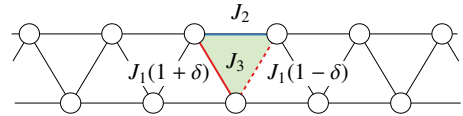}
 \caption{(Color online) A schematic figure of the $J_1$-$J_2$-$J_3$ model with the
 bond alternation $\delta$ \eqref{eq:dJ1J2J3}.
 Alternating nearest-neighbor interactions $J_1(1\pm \delta)$ are
 depicted by the red solid  and dashed lines.
 The next-nearest interaction $J_2$ is represented by the blue line.
 The three-spin interaction $J_3$ works on a triangle (green shaded
 area). 
 }
 \label{fig:dJ1J2J3}
\end{figure}

Recently, Michaud \textit{et al.} discussed a special case of this
model with  $J_2=\delta=0$.~\cite{Michaud-VMM}
They have shown that this ``$J_1$-$J_3$''
model is reduced to the MG model for $S = 1/2$, and that a
doubly degenerate dimerized state appears as the exact ground
state when the ratio of the couplings is $J_3/J_1 = 1/[4S(S+1)-2]$.
In this sense, one may regard the $J_1$-$J_3$ model as a
generalization of the MG model.
Michaud \textit{et al.} argued that the $J_1$-$J_3$ model has exact dimerized ground states
for \emph{any} $S$.
While their conclusion is undoubtedly interesting, the mechanism
that allows the $J_1$-$J_3$ model to possess an exact ground state is still unclear.
In this paper, we prove rigorously that the fully dimerized state is the exact ground state of
the $J_1$-$J_3$ model from a more comprehensive point of view.
Namely, we present a proof of the fact that the general extension of
the $J_1$-$J_3$ model, that is, the ``$J_1$-$J_2$-$J_3$'' model with bond
alternation \eqref{eq:dJ1J2J3}, has an exact dimerized ground state.
Our proof, besides its rigorousness, gives a physical answer to the natural
question of why an exact ground state is feasible in the general model \eqref{eq:dJ1J2J3}.
\begin{figure}[t!]
 \centering
 \includegraphics[width=70mm]{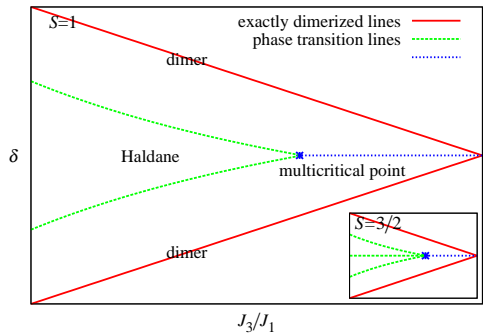}
\caption{(Color online) Schematic ground state phase diagram for the model
\eqref{eq:dJ1J2J3} with $J_2=0$ for $S=1$ and $3/2$.  The solid lines
correspond to the exact dimer ground-state regions.  The dashed lines
depict the critical lines between several dimerized phases including the
Haldane phase.  These critical lines are expected to be of the $c=1$ Gaussian
type.  The stars denote multicritical points where the critical lines
merge as described by $\mathrm{SU}(2)_{2S}$ WZW theory with $c=3S/(S+1)$.}
\label{fig:ipd}
\end{figure}

One noteworthy feature of the model \eqref{eq:dJ1J2J3}
is that it contains experimentally feasible interactions only.~\cite{Michaud-VMM}
This feature will be relevant to experimental studies directly or indirectly.
Thus, it will be beneficial for experimental studies to investigate the
spin chain \eqref{eq:dJ1J2J3} away from the exactly dimerized regions.
For simplicity, we consider the phase diagram of Eq.~\eqref{eq:dJ1J2J3}  for $J_2=0$ 
in the $\delta$-$J_3/J_1$ space.
All the physical essence is encoded in this limited parameter region.
Since the exact ground states are in the dimerized phase,
there must be phase transition lines between the dimerized regions
 and the origin of the parameter space which corresponds to the pure Heisenberg model
(i.e., $J_2=J_3=\delta=0$).
As discussed by Michaud {\it et al.}~\cite{Michaud-VMM,Michaud-MM}, the
phase transition occurs at a point $J_{3c}/J_1 >0$ with $\delta=0$ and
it is expected that the level-$2S$ $\mathrm{SU}(2)$ Wess-Zumino-Witten
theory [$\mathrm{SU}(2)_{2S}$ WZW theory] is realized there.
This means that various relevant interactions are canceled at the critical point.
On the other hand, for a fixed $J_3/J_1<J_{3c}/J_1$,
there should be $2S$ successive dimerization transitions between
$2S+1$ phases with different VBS configurations when $\delta$ is
varied from $\delta=-1$ to $\delta=1$.
These transitions are expected to be of Gaussian type.
Therefore, the ground-state phase diagram
of the general-$S$ model \eqref{eq:dJ1J2J3} will consist of phase transitions of
different universality classes. A field-theoretical point of view
allows a comprehensive understanding of such complicated situations. 
The field-theoretical description also enables us to
perform an accurate numerical analysis of the phase diagram
by the level-spectroscopy method.~\cite{Kitazawa}
We will determine the
phase diagram for the $S=1,3/2$ and $J_2=0$ systems of Eq.~\eqref{eq:dJ1J2J3}
numerically as shown in Fig.~\ref{fig:ipd}.

The rest of this paper is organized as follows:
In Sec.~\ref{sec:proof}, we prove that there exist regions of the exact dimerized ground
state of the spin-$S$ AFH chain (\ref{eq:dJ1J2J3})
in the $(\delta,J_1,J_2,J_3)$ parameter space.
In Sec.~\ref{sec:pd}, we discuss the universality class of the phase
transition of thismodel with $J_2=0$ based on a field-theoretical argument.
Furthermore, we determine the phase diagrams of this model
for  $S=1$ and $S=3/2$ cases numerically,
using exact diagonalization and the level-spectroscopy method.
In the Appendix and Supplemental Material,~\cite{SM} the detailed
calculation for the proof of the exact ground state will be presented.

\section{Proof of the exact dimerized ground states}
\label{sec:proof}

In this section, we prove that Eq.~\eqref{eq:dJ1J2J3} has parameter regions
of exact dimer ground states for general $S$.
First, let us reconsider the Majumdar-Ghosh model [Eq.~\eqref{eq:dJ1J2J3} with
$S=1/2$, $\delta=J_3=0$, and $J_2=J_1/2$],
\begin{equation}
 \mathcal{H}_{\rm MG}=
  J_1\sum_i
 (\bm{S}_i \cdot \bm{S}_{i+1}
 +{\textstyle\frac{1}{2}}\bm{S}_{i-1} \cdot \bm{S}_{i+1}).
 \label{eq:J1J2}
\end{equation}
This model has two fully dimerized states as the exact ground state,
\begin{equation}
 \begin{split}
  \ket{\Psi_{\mathrm{odd}}}&=\prod_{i\in \mathbb Z} \ket{S(2i-1,2i)},\\
  \ket{\Psi_{\mathrm{even}}}&=\prod_{i\in \mathbb Z} \ket{S(2i,2i+1)},
 \end{split}
\label{eq:gs}
\end{equation}
where $\ket{S(i,i+1)}$ denotes the singlet state formed by the spins
at sites $i$ and $i+1$. This can easily be proven if we rewrite
Eq.~(\ref{eq:J1J2}) using the following projection operator:
\begin{equation}
 Q_{i}=(\bm{S}_{i-1}+\bm{S}_{i}+\bm{S}_{i+1})^2-S(S+1).
  \label{opQ}
\end{equation}
For the singlet states, the three spins form an $S=1/2$
composite spin, so that $\braket{Q}_i=0$
while $\braket{Q}_i\geq 0$ for other states.
However, for $S \ge 1$, the singlet states (\ref{eq:gs}) are no longer
the ground states of this projection operator.
\begin{figure}[b!]
 \centering
 \includegraphics[width=80mm]{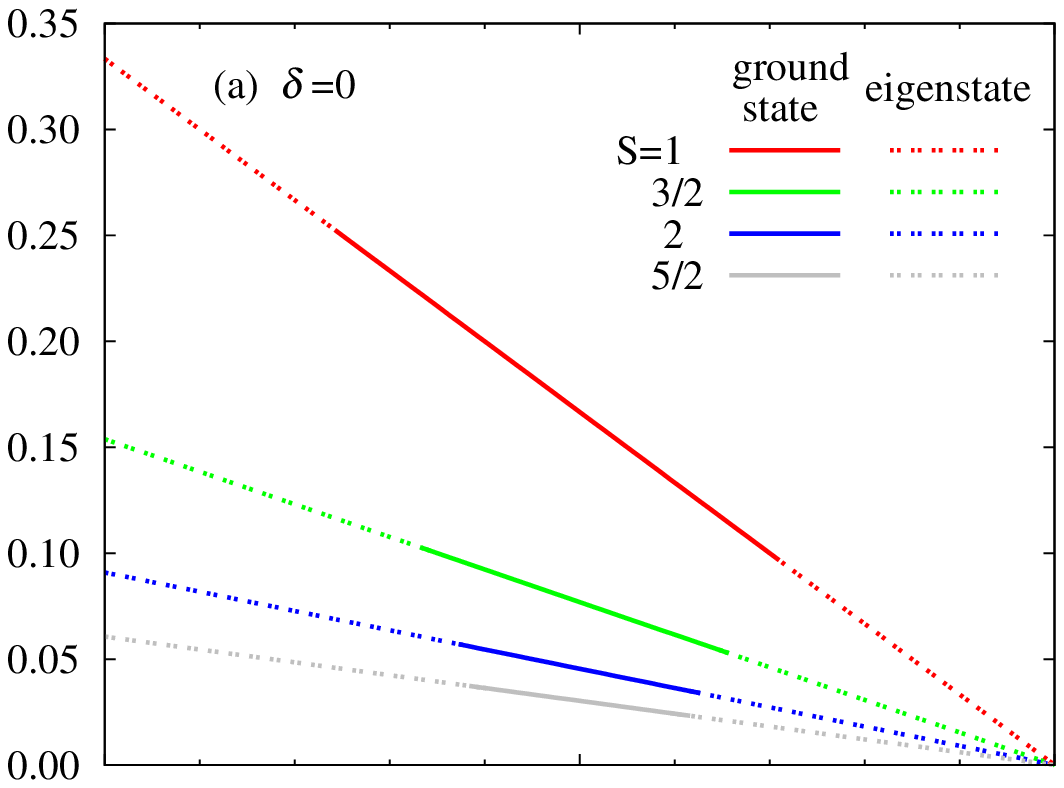}
 \includegraphics[width=80mm]{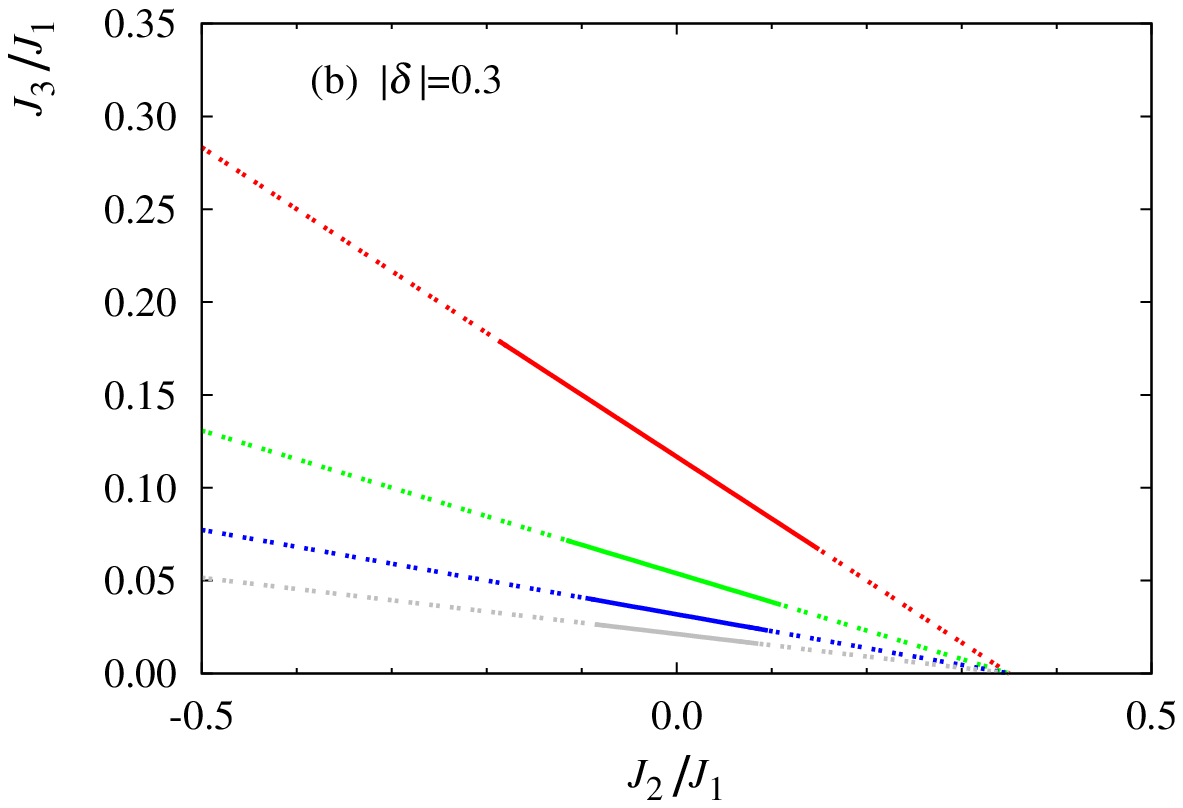}
 \caption{(Color online) The parameter space of the model \eqref{eq:dJ1J2J3} for
 several $S\geq 1$ with (a) $\delta=0$ and (b) $|\delta|=0.3$. The solid
 (dotted) lines denote the regions where the dimerized state is the
 exact ground state (eigenstate).}
 \label{fig:egs}
\end{figure}

Next, we consider the $J_1$-$J_3$ model [Eq.~\eqref{eq:dJ1J2J3} with
$J_2=\delta=0$] which was discussed by Michaud {\it et
al.}:~\cite{Michaud-VMM}
\begin{align}
 \mathcal H_{J_1\mathchar`-J_3}
 &= J_1 \sum_j \bm S_j \cdot \bm S_{j+1} \notag \\
 &+J_3\sum_i 
 \bigl[ (\bm S_{i-1} \cdot  \bm S_i)(\bm S_i \cdot \bm S_{i+1})
 +\mathrm{H.c.}\bigr].
\end{align}
This Hamiltonian for $J_3/J_1 = 1/[4S(S+1)-2]$ can
be written as
\begin{equation}
 \mathcal{H}_{J_1\mathchar`-J_3}
  =\frac{J_1}{16S(S+1)-8}\sum_i P_i - \frac{J_1}{2}NS(S+1),
  \label{eq:tH_J1J3}
\end{equation}
where $N$ denotes the total number of the spins, and $P_i$ is a
local projection operator defined by
\begin{equation}
\begin{split}
P_i
=&(\bm{S}_{i-1}+\bm{S}_i)\cdot
 \left[(\bm{S}_i+\bm{S}_{i+1})^2-1\right]
 \cdot(\bm{S}_{i-1}+\bm{S}_i)\\
=&(\bm{S}_i+\bm{S}_{i+1})\cdot
 \left[(\bm{S}_{i-1}+\bm{S}_i)^2-1\right]
 \cdot(\bm{S}_i+\bm{S}_{i+1}).
\end{split}
\label{opP}
\end{equation}
It follows from $S_i^{\alpha}\ket{S(i,j)}=-S_j^{\alpha}\ket{S(i,j)}$
($\alpha=x,y,z$) that $P_i$ projects out (i.e., annihilate) the singlet
states, 
\begin{equation}
 P_i\ket{S(i,i+1)} = P_i\ket{S(i-1,i)} = 0.
\end{equation}
Therefore, Eq.~(\ref{eq:gs}) is
an exact {\it eigenstate} of the $J_1$-$J_3$ model.~\cite{Michaud-VMM}

In order to show that Eq.~(\ref{eq:gs}) is the exact {\it ground state},
we have to prove that $P_i$ is positive semidefinite.  For this purpose,
let us introduce the following two local projection operators:
\begin{equation}
 R_{i} \equiv \frac{1}{2}P_{i}+(S+1)Q_{i},
  \quad
  R'_{i} \equiv \frac{1}{2}P_{i}-SQ_{i}.
  \label{eq:R_def}
\end{equation}
Since $[P_i,Q_i]=0$ and also $[R_i,R'_i]=0$, $R_i$ and $R'_i$ project
out $\ket{S(i-1,i)}$ and $\ket{S(i,i+1)}$.  Here one can show that $R_i$
and $R'_i$ are positive semidefinite, while $Q_{i}$ is not. 
Let $\ket{\psi}$ be an eigenstate of $R_i$ and $R'_i$ simultaneously.
We can confirm the non-negativity of its eigenvalues
without making any {\it ad hoc} assumption:
\begin{equation}
 \bra{\psi}R_i\ket{\psi} \ge 0, \quad \bra{\psi}R'_i\ket{\psi} \ge 0,
  \label{eq:eigenvalues}
\end{equation}
where the equalities are satisfied for the singlet states.  The relations
\eqref{eq:eigenvalues} are proven by expanding the three-sites wave
functions explicitely into Ising-like bases (see the Appendix).
Then $P_i$ is also proven to be positive semidefinite
due to the relation
\begin{equation}
 P_i=\frac{S}{2S+1}R_i+\frac{S+1}{2S+1}R'_i.
  \label{conPR}
\end{equation}
Thus the $J_1$-$J_3$ model has the exact doubly degenerate dimerized
ground state.
In terms of $R_i$ and $R'_i$, it is obvious that $Q_i$ is not positive
semidefinite:
\begin{equation}
 Q_i = \frac 1{2S+1} (R_i-R'_i).
  \label{eq:Q}
\end{equation}
Thus, the naive extension of the MG model $\mathcal H_{\rm MG} =
(J_1/2)\sum_i Q_i +\mathrm{const}$ to the general-$S$ case fails to have
an exact dimerized ground state.

The above argument enables us to generalize the $J_1$-$J_3$ model to
Eq.~\eqref{eq:dJ1J2J3} with $\delta=0$, by constructing the Hamiltonian
in the following form:
\begin{equation}
 \tilde{\mathcal H}_{J_1\mathchar`-J_2\mathchar`-J_3}
  =\sum_i \left( A R_i +B R'_i \right),\quad A, B \ge 0,
  \label{eq:H_AB}
\end{equation}
where $\tilde{\mathcal{H}}_{J_1\mathchar`-J_2\mathchar`-J_3}\equiv
\mathcal{H}_{J_1\mathchar`-J_2\mathchar`-J_3}-E_0$ with $E_0$ being the
ground state energy of
$\mathcal{H}_{J_1\mathchar`-J_2\mathchar`-J_3}$. Furthermore, the model
can be extended to $\delta\neq 0$ cases as
\begin{equation}
 \mathcal{H}_{\delta\mathchar`-J_1\mathchar`-J_2\mathchar`-J_3}
  = \mathcal H_{J_1\mathchar`-J_2\mathchar`-J_3}
  +\sum_{i \in I_\delta}
  \delta' (\bm{S}_i+\bm{S}_{i+1})^2,
  \label{eq:H_dJ1J2J3_exact}
\end{equation}
where $I_\delta$ stands for a set of odd (even) integers for $\delta>0$
($\delta<0$). $\delta'>0$ and other parameters are related as
\begin{align}
& \delta'=\frac{1+\delta}{1-\delta}
  \left[
   A(2S^2+3S)+B(2S^2+S-1)
  \right],
 \label{eq:exact_delta} \\
&\frac{J_3}{(1-| \delta |)J_1-2J_2} = \frac 1{4S(S+1)-2}.
\label{eq:exact}
\end{align}
Since the operator $(\bm S_i + \bm S_{i+1})^2$ is obviously positive
semidefinite and projects out the singlet state $\ket{S(i,i+1)}$,
a nonzero $\delta$ lifts the doublet degeneracy of
$\ket{\psi_{\mathrm{odd}}}$ and $\ket{\psi_{\mathrm{even}}}$.  For
$\delta>0$ ($\delta<0$), the fully dimerized state
$\ket{\psi_{\mathrm{odd}}}$ ($\ket{\psi_\mathrm{even}}$) is the
\emph{unique} ground state of the model \eqref{eq:H_dJ1J2J3_exact}.
The parameter regions of the exact ground state given by
Eq.~\eqref{eq:exact} are shown in Fig.~\ref{fig:egs}.

The point of our proof lies in the finding that the operator $P_i$ in
Eq.~\eqref{eq:tH_J1J3} can be written as two positive semidefinite operators
$R_i$ and $R'_i$, both of which project out the singlet states
$\ket{S(i-1,i)}$ and $\ket{S(i,i+1)}$.  Our proof clarified that the
operators $R_i$ and $R'_i$ that project out the singlet states are
essential for realizing the exact dimerized ground state.

\section{Phase diagram of $J_2=0$ chains}\label{sec:pd}

In this section, we discuss the phase diagrams of the model
\eqref{eq:dJ1J2J3} using a field-theoretical approach. For simplicity we
consider only $J_2=0$ cases in the $\delta$-$J_3/J_1$ space.
The phase diagram of a related model including the bond alternation and
the next-nearest-neighbor interaction $J_2$ is investigated in detail in
Ref.~\onlinecite{Takayoshi}.

\subsection{\label{sec:wzw}Effective field theories}

We introduce effective field theories of the AFH chain
\eqref{eq:dJ1J2J3} for $J_2=0$, such as the $\mathrm O(4)$ and $\mathrm
O(3)$ nonlinear $\sigma$ models (NLSMs).  All the field theories are derived
from the $SU(2)_{k}$ WZW theory.
These field theories gives basic languages needed to understand the static and
dynamics of this system.

\subsubsection{The multicritical point}\label{sec:multi}

At a point $(\delta ,J_3) = (0,J_{3c})$ with $J_{3c}>0$, the Gaussian
critical lines merges.  This point is the multicritical point and the
semi-infinite line $J_3 > J_{3c}$ on the horizontal axis is the
first-order transition line.  Low-energy physics at the multicritical
point is described by the $\mathrm{SU(2)}_{2S}$ WZW theory and the
central charge $c =3S/(1+S)$.~\cite{Michaud-VMM} The
$\mathrm{SU(2)}_{2S}$ (WZW) theory has the following Euclidean action,
\begin{align}
 \mathcal A_{2S}
 &= \frac S{8\pi}\int_{S^2}d^2x \Tr (\partial_\mu U \partial_\mu U^{-1})
 + 2S\Gamma_{\mathrm{WZ}}.
\end{align}
$U\in \mathrm{SU}(2)$ is a primary field of the $\mathrm{SU}(2)_{2S}$
WZW theory.  The second term is the Wess-Zumino term,
\begin{equation}
\begin{split}
 \Gamma_{\mathrm{WZ}}=&
 -\frac{i}{24\pi}\int_{B^3} d^3x \, \epsilon^{\alpha \beta \gamma} \mathrm{Tr}
 (U^{-1}\partial_\alpha U\\
&\times U^{-1}\partial_\beta U U^{-1} \partial_\gamma U).
\end{split}
\end{equation}
The two-dimensional space $\{(\tau,x)|\tau, x\in \mathbb R^2\}$ is
identified as the two-dimensional sphere $S^2$ by a one-point
compactification of the infinity.  $B^3$ is a three-dimensional ball
whose boundary is $\partial B^3 = S^2$.  The primary field $U \in
\mathrm{SU}(2)$ can be represented by a $2\times 2$ matrix with four
real parameters $\phi_\alpha$ ($\alpha = 0,1,2,3$),
\begin{align}
 U 
 &= \phi_0 I+ i\bm \phi \cdot \bm \sigma.
 \label{eq:U}
\end{align}
$I=\mathrm{diag}(1,1)$ is the $2\times 2$ unit matrix.
$\bm \sigma = (\sigma_1, \sigma_2, \sigma_3)$ denotes the Pauli
matrices.
The parameters $\phi_\alpha$ are constrained to the three-dimensional
sphere $S^3$,
\begin{equation}
 \sum_{\alpha=0}^3 {\phi_\alpha}^2 = {\phi_0}^2 + \bm \phi^2 = 1.
  \label{eq:const}
\end{equation}
Therefore, the $\mathrm{SU(2)}_{2S}$ WZW theory is equivalent to the $O(4)$ NLSM
\begin{equation}
 \mathcal A_{2S} = \frac S{4\pi}\int d\tau dx \sum_{\alpha = 0}^3
  (\partial_\mu \phi_\alpha)^2 + 2S\Gamma_{\mathrm{WZ}}.
  \label{eq:O4NLSM}
\end{equation}
The contraction $(\partial_\mu \phi_\alpha)^2 = (\partial_\tau
\phi_\alpha)^2 + (\partial_x\phi_\alpha)^2$ is assumed.  Whereas a
$(2+1)$-dimensional version of the $O(4)$ NLSM \eqref{eq:O4NLSM} is
fully exploited to understand the deconfined quantum criticality of
two-dimensional quantum
antiferromagnets,~\cite{senthil_science,tanaka_NLSM} the
$(1+1)$-dimensional version \eqref{eq:O4NLSM} is not used in
one-dimensional quantum spin systems.  In the following,
we will see that the representation \eqref{eq:O4NLSM} also gives
insight on the problem of one-dimensional quantum antiferromagnets.
$\phi_0$ and $\bm \phi$ represent order parameters of the dimerization
and the Neel order, respectively:
\begin{equation}
 \begin{pmatrix}
  S(S+1)\phi_0(x) \\
  \sqrt{S(S+1)}\bm \phi(x)
 \end{pmatrix}
 \simeq
 \begin{pmatrix}
  (-1)^j \bm S_j \cdot \bm S_{j+1} \\
  (-1)^j \bm S_j
 \end{pmatrix}.
  \label{eq:phi_unit}
\end{equation}
The $\mathrm{SO}(4)$ symmetry of the WZW theory is easily broken
by relevant interactions. 
Indeed, the $\mathrm{SO}(4)$ group is  homomorphic to
$\mathrm{SU}(2)_L \times \mathrm{SU}(2)_R$.
The subscripts $L$ and $R$ denote the holomorphic (right-moving) and
the antiholomorphic (left-moving) degrees of freedom.
The manifestation of the $\mathrm{SO}(4)$ symmetry 
means the decoupling of the holomorphic and
antiholomorphic parts  of the effective field theory, which
is nothing but the fixed point theory \eqref{eq:O4NLSM}.
For instance, the most relevant interaction in the WZW theory is
$\mathrm{Tr} U = 2\phi_0$.
It apparently lowers the $\mathrm{SO}(4)$ symmetry
to $\mathrm{SO}(3)$ if $\mathrm{Tr} U$ acquires a nonzero expectation
value as in the dimerized phase.

The $\mathrm{SO}(4)$ symmetry is naturally manifested in the lattice model
\eqref{eq:dJ1J2J3}.  Let us roughly determine a point where the
$\mathrm{SO}(4)$ symmetry is realized in the phase diagram.  First of
all, it must be on the horizontal axis, $\delta = 0$, because the bond
alternation $(-1)^j \bm S_j \cdot \bm S_{j+1} \sim S(S+1)\phi_0$ cannot
respect the $\mathrm{SO}(4)$ symmetry.  The Hamiltonian
\eqref{eq:dJ1J2J3} with $J_2 = 0$ and $\delta = 0$ is written in terms
of the $(\phi_0, \bm \phi)$ fields as follows,
\begin{align}
 \mathcal H
  &\sim \int \frac{dx}a \, \Bigl[-J_1S(S+1) \bm \phi(x) \cdot \bm \phi(x+a)
 \notag \\ & \qquad 
 -J_3\bigl[S(S+1)\bigr]^2\phi_0(x) \phi_0(x+a) + \cdots \Bigr].
 \label{eq:H_phi}
\end{align}
$a$ is the lattice spacing of the spin chain.
The $\mathrm{SO}(4)$ symmetry exists only when
the right hand side of Eq.~\eqref{eq:H_phi}
takes the form of $-JS(S+1)\sum_{\alpha=0}^3 \phi_\alpha(x)
\phi_\alpha(x+a) + \cdots$.

That is, the $\mathrm{SO}(4)$ symmetry is realized in the spin chain
\eqref{eq:dJ1J2J3} at
\begin{equation}
 \biggl(\delta, \frac{J_{3c}}{J_1}  \biggr)
  = \biggl(0,\frac{C_S}{S(S+1)} \biggr).
  \label{eq:O4}
\end{equation}
$C_S$ is a nonuniversal constant which may depend on the spin quantum
number $S$.  It is generally difficult to determine the precise value of $C_S$.
But, one can speculate that the
constant $C_S$ satisfies $0< C_S \lesssim \frac 14$ because it lies
between the origin $(0,0)$ and the exactly dimerized point, $(\frac
{1}{4S(S+1)-2}, 0)$.  In fact, $J_{3c}/J_1 = 0.111$ in the $S=1$ AFH
chain \eqref{eq:dJ1J2J3} leads to $C_{S=1}=0.222$, which satisfies $0 <
C_{S=1} < \frac {1}{4}$.

The effective field theory of the AFH chain \eqref{eq:dJ1J2J3} in
the vicinity of the multicritical point has the following Euclidean action:
\begin{equation}
 \mathcal{A} 
 \simeq \mathcal{A}_{2S} - \int d\tau dx \, \bigl[ c_\delta J_1\delta \Tr U
 +c_3 (J_3-J_{3c}) (\Tr U)^2
 \bigr]
 \label{eq:A}
\end{equation}
$c_\delta$ and $c_3$ are positive dimensionless constant.
Every effective field theory in distinctive regions of the
phase diagram originates from the perturbed $\mathrm{SO}(2)_{2S}$ WZW theory
\eqref{eq:A}.

The qualitative effects of the perturbations are easily seen as follows.
When we move away from the multicritical point vertically ($J_3 =
J_{3c}$), the $\phi_0$ field is pinned to either of $\phi_0 = \pm 1$ in
order to optimize the potential $\Tr U = 2\phi_0$.  The pinning value
depends on the sign of $\delta$.  But, irrespective of the sign, the
spin chain \eqref{eq:dJ1J2J3} is drawn into the fully dimerized phase.
On the other hand, when we move away horizontally ($\delta = 0$), the
$\phi_0$ field is pinned to $\pm 1$ if $J_3 > J_{3c}$ and $0$ if $J_3 <
J_{3c}$.  Only the former case corresponds to the fully dimerized phase
where the one-site translational symmetry is spontaneously broken.

\subsubsection{Between the  multicritical point and the origin}
The perturbed WZW theory \eqref{eq:A} with $\delta = 0$ and $J_3<
J_{3c}$ is equivalent to the $\mathrm O(3)$ NLSM,~\cite{AffleckHaldane}
\begin{equation}
 \mathcal A = \int d\tau dx \, \frac v{2g}(\partial_\mu \bm n)^2 +
  i\Theta Q
  \label{eq:A_O3}
\end{equation}
 with a topological angle $\Theta = 2\pi S$.
$\bm n = (n^x, n^y, n^z)$ is a three-dimensional unit vector.
The $\bm n$ field is related to the original spin variable $\bm S_j$ via
\begin{equation}
 \bm S_j \simeq (-1)^j S \bm n(x) \sqrt{1-\biggl( \frac{\bm
  L(x)}S\biggr)^2}   + \bm L(x).
  \label{eq:S2nL}
\end{equation}
Each component of $\bm n(x)$ represents a triplet one-magnon excitation
near the wavenumber $q=\pi/a$, and $\bm L(x) = \frac 1{gv}\bm n \times
\partial_t \bm n$ describes two-magnon excitations near $q=0$.  The term
$i\Theta Q$ in Eq.~\eqref{eq:A_O3} is the so-called $\Theta$ term.  $g=2/S$
is a coupling constant.  $Q$ is an integer, which represents a winding
number of the configuration of $\bm n$ field on the two-dimensional
sphere $S^2$.  As Haldane first proposed,~\cite{haldane_1983b} the
$\mathrm O(3)$ NLSM \eqref{eq:A_O3} describes a gapped spin chain when
$S$ is an integer.

\subsubsection{On the critical line}
Bond alternation modifies the coupling of the $\Theta$ term,~\cite{kato_BAHAF}
\begin{equation}
 \Theta = 2\pi S(1-\delta).
\end{equation}
By increasing $|\delta|$ from zero, we can change the coupling $\Theta$ from
$\Theta \equiv 0$ (mod $2\pi$) to $\Theta \equiv \pi$ (mod $2\pi$) in
the case of integer $S$, and from $\Theta \equiv \pi$ (mod $2\pi$) to
$\Theta = 0$ (mod $2\pi$) in the case of half-integer $S$.
The $\mathrm O(3)$ NLSM \eqref{eq:A_O3} with $\Theta \equiv \pi$ (mod $2\pi$) is
equivalent to the massless free-boson theory,
\begin{equation}
 \mathcal A = \frac 12 \int d\tau dx \,(\partial_\mu
  \varphi)^2.
  \label{eq:A_G}
\end{equation}
The boson field $\varphi$ is compactified as  $\varphi \sim \varphi +
\sqrt{2\pi}$.
For $S=1$,
the critical theory \eqref{eq:A_G} describes the low-energy physics
exactly on the $c=1$ Gaussian critical line.
This will also be the case for the $S = 3/2$ case.

\subsection{Level spectroscopy}\label{sec:numerical}
\begin{figure}[t]
 \centering
 \includegraphics[width=80mm]{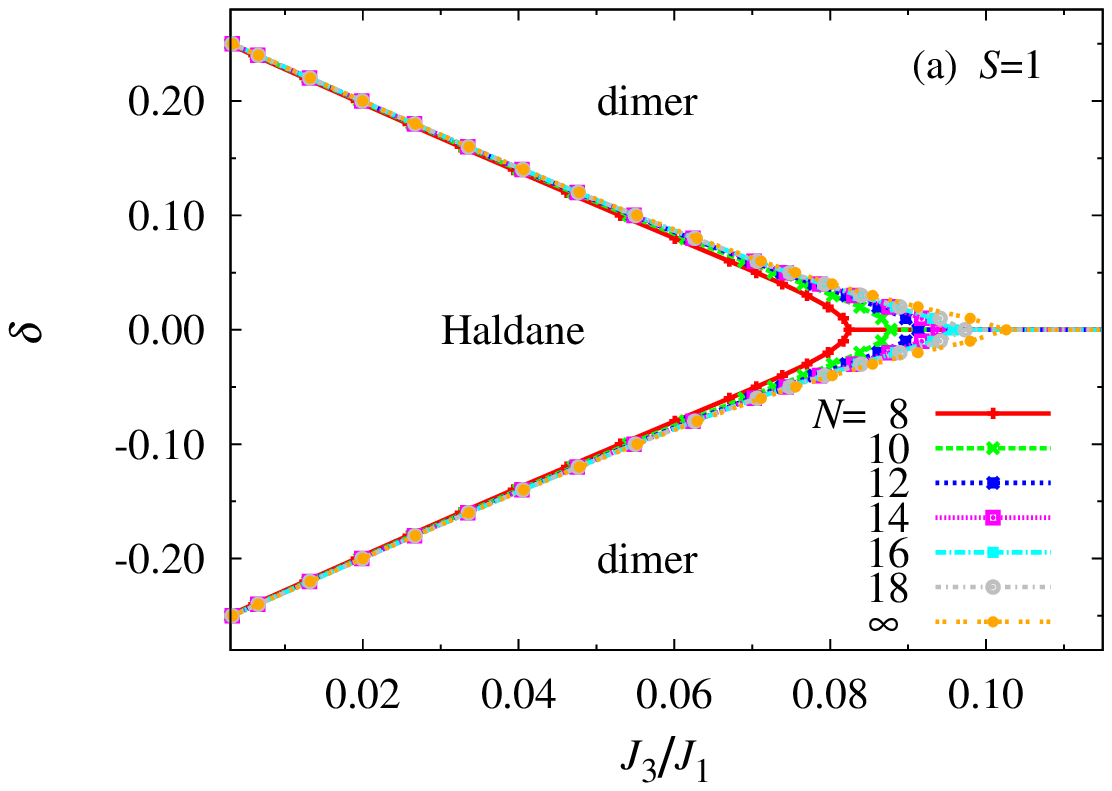}
 \includegraphics[width=80mm]{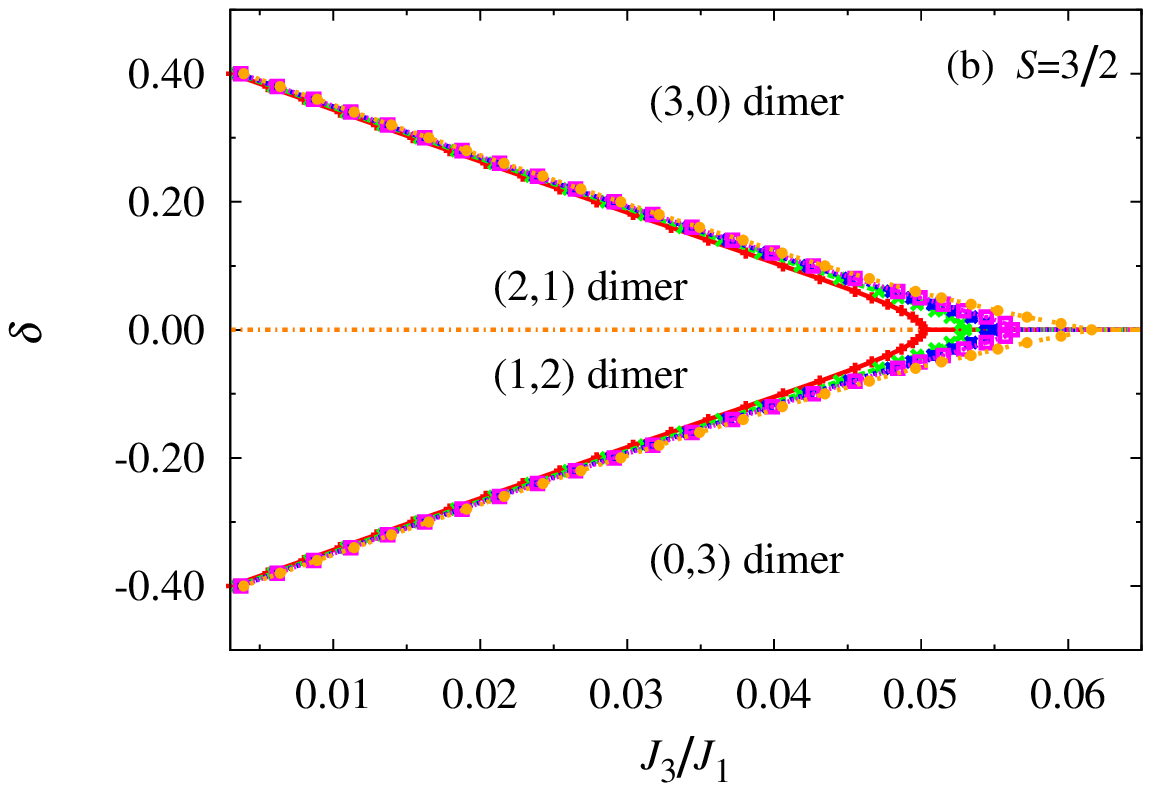}
 \caption{(Color online) Ground-state phase diagram of the bond-alternating Heisenberg
 model with three-spin interactions [Eq.~(\ref{eq:dJ1J2J3}) with $J_2=0$]
 for (a) $S=1$ and (b) $S=3/2$. The multiciritical points are located at
 $(J_3/J_1)_c=0.11$ for $S=1$ and at $(J_3/J_1)_c=0.062$ for $S=3/2$.}
 \label{fig:PD}
\end{figure}

The phase diagrams of the $J_2=0$ chains for $S=1$ and
$S=3/2$ are determined by level spectroscopy which identifies
the phase transition as a level-crossing point of the lowest
excitation levels obtained by exact diagonalization. The finitesize
effect in this method is very small, since the logarithmic
corrections are canceled at the level-crossing point, and only
the power dependence of the size effect remains. The present
Haldane-dimer and dimer-dimer transitions can be determined
by level crossings of the lowest two excitations with different
parities under antiperiodic boundary conditions.~\cite{Kitazawa}
These level crossings can also be interpreted as points of change of the
$(m,n)$-type VBS configurations, which are defined as
\begin{equation}
 |\Psi_{\rm VBS}^{(m,n)}\rangle
  \equiv
\prod_{k=1}^{N/2}
  (B_{2k-1,2k}^{\dag})^{m}(B_{2k,2k+1}^{\dag})^{n}|{\rm vac}\rangle,
  \label{eqn:VBS}
\end{equation}
where $B_{i,j}^{\dag}\equiv
a_{i}^{\dag}b_{j}^{\dag}-b_{i}^{\dag}a_{j}^{\dag}$ and $|{\rm
vac}\rangle$ is the vacuum with respect to bosons.~\cite{Arovas} Here,
the Schwinger boson operator $a_j^{\dag}$ ($b_j^{\dag}$) increases the
number of up (down) $S=1/2$ variables under symmetrization. The integers
$m$ and $n$ satisfy $m+n=2S$.
The level-crossing points also correspond to zero points of the
expectation values of the twist order parameters.\cite{Nakamura-T}
Moreover, for $S=1$the two excitation spectra for the level crossing
can be related to the two different types of string order
parameters.~\cite{Nakamura}

The phase diagrams obtained by this method are shown in
Fig.~\ref{fig:PD}.  The extrapolation of the numerical data has been
done assuming the function $\delta_c(N)=\delta_c(\infty)+A/N^2+B/N^4$.
The finite-size effect becomes larger for small-$\delta$ regions, but
the extrapolated values [$(J_3/J_1)_c=0.11$ for $S=1$,
$(J_3/J_1)_c=0.062$ for $S=3/2$] well agrees with the ones obtained in
Refs.~\onlinecite{Michaud-VMM,Michaud-MM}.
The global structure of the phase diagram Fig.~\ref{fig:PD} is
similar to that for the $S = 1$ chain with a bilinear-biquadratic
interaction. This is due to the fact that the three-spin interaction
and the bilinear-biquadratic interaction will take almost the
same form in the continuum limit.~\cite{Kitazawa-N}
\begin{figure}[t!]
 \centering
 \includegraphics[width=80mm]{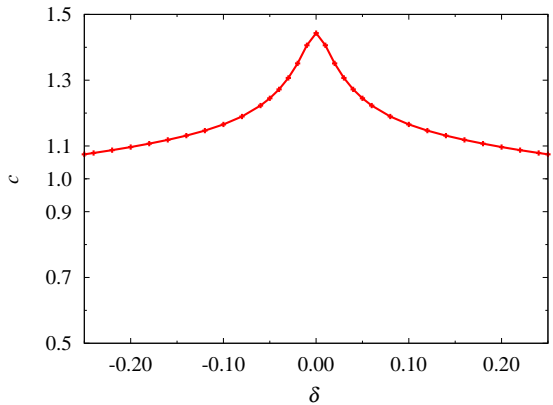}
 \caption{(Color online) Conformal charge on the Haldane-dimer transition line in $S=1$
 chain obtained from the size-scaling of the ground state
 $N=8$-$18$. The value approaches to $c=1$ ($c=3/2$) for $\delta\neq 0$
 ($\delta=0$).}
 \label{fig:CC}
\end{figure}

We calculate the conformal charge $c$ along the critical line for the
$S=1$ system using the standard finite-size scaling of the ground state,
\begin{equation}
 E_0(N)=\varepsilon_0 N-\frac{\pi v}{6N}c,
\end{equation}
where the spin-wave velocity is calculated as
\begin{equation}
 v=\lim_{N\to\infty}\frac{E(N,S=1,k=4\pi/N)-E_0(N)}{4\pi/N}.
\end{equation}
As shown in Fig.~\ref{fig:CC}, although the finite-size effect remains
large, $c\to 1.5=3S/(S+1)$ around the multicritical point $\delta=0$
whereas $c\to 1$ for $\delta\neq 0$ as was predicted by the field-theoretical arguments.

\section{Summary}\label{sec:summary}

We have discussed the dimerizations of the spin-$S$ AFH chains with
three-spin interactions and bond alternations.
We have rigorously proven that the AFH chains can possess exact
dimerized ground states: Two local projection operators
$R_i$ and $R'_i$ defined in Eqs.~(\ref{opR}) and (\ref{opRd})are introduced and
the Hamiltonians with exactly dimerized ground states are
written as a linear combination of them.
Then the exact ground states turns out to
lie on the surface $\frac{J_3}{(1-\delta)J_1-2J_2} = \frac
1{4S(S+1)-2}$. This argument is considered to be a generalization of the
Majumdar-Ghosh model to arbitral $S$. It is also possible to construct
higher-dimensonal version of the present model, such as the
Shastry-Sutherland model.

We have further discussed the ground-state phase diagram of the models whose
Hamiltonian is defined by \eqref{eq:dJ1J2J3} by introducing several
effective field theories.  At the multicritical point low-energy physics
is described by the $\mathrm{O}(4)$ nonlinear $\sigma$ model.
However, on the phase transition lines except at the multicritical point,
the low-energy physics is described by $\mathrm{O}(3)$ nonlinear $\sigma$ model.
We have also obtained the phase diagram for $S=1$ and $S=3/2$ chains by the
level-spectroscopy method, and confirmed that the central charge on the
critical lines changes rapidly from $3S/(S+1)$ to $1$ when the bond
alternation increases.

\section*{acknowledgments}
The authors are grateful to Masaki Oshikawa for fruitful discussions.
S. C. F.  is supported by the Swiss National Foundation under
MaNEP and Division II.
M. N. acknowledges support from MEXT Grant-in-Aid No.23540362.

\appendix*

\section{Positivity of $R_i$ and $R'_i$}\label{sec:psp}

We prove positivity of the operators $R_i$ and $R'_i$ defined in
Eqs.~(\ref{opR}) and (\ref{opRd}) for arbitrary simultaneous eigenstate
$\ket{\psi}$.  Indeed, one can take such a state, because
$[R_2,R'_2]=0$.  Thanks to the translational symmetry of the $J_1$-$J_3$
model, we may assume $i=2$ without loss of generality.  Since $R_i$ and
$R'_i$ are involved with three neighbouring spins $\bm S_{i-1}$, $\bm
S_i$ and $\bm S_{i+1}$, one only need to a three-site subspace, in which
an arbitrary state $\ket{\psi}$ can be written as
\begin{equation}
\label{DefStat}
\ket{\psi}=\sum_{m_1, m_2, m_3} C_{m_1,m_2, m_3} \ket{m_1,m_2,m_3},
\end{equation}
with
\begin{equation}
\sum_{m_1,m_2,m_3} |C_{m_1,m_2,m_3}|^2=1,
\end{equation}
where $m_j \, (j=1,2,3)$ is the $S^z$ quantum number of each spin on
site $j$ and thus $|m_j| \le S$.

After straightforward calculations,\cite{SM} one finds that the
expectation values of the state $\ket{\psi}$ are non-negative,
\begin{align}
\!\bra{\psi}\!R_2\!\ket{\psi}
&\!=\sum_{m_1,m_2,m_3}\!
\big|
\alpha C_{m_1,m_2,m_3}
\!+\!\beta C_{m_1-1,m_2+1,m_3}\notag
\\
&\;+\!\gamma C_{m_1,m_2-1,m_3+1}
\!+\!\theta C_{m_1-1,m_2,m_3+1}
\big|^2,
\label{opR}\\
\!\bra{\psi}\!R'_{2}\!\ket{\psi}
&\!=\sum_{m_1,m_2,m_3}\!
\big|
\alpha' C_{m_1,m_2,m_3}
\!+\!\beta' C_{m_1-1,m_2+1,m_3} \notag
\\
&\;+\!\gamma' C_{m_1,m_2\!-\!1,m_3+1}
\!+\!\theta' C_{m_1\!-\!1,m_2,m_3+1} 
\big|^2,
\label{opRd}
\end{align}
where
\begin{equation}
\vspace{+2mm}
\begin{split}
\alpha\!=\!&\sqrt{\!(S\!+\!m_1)(S\!+\!m_2\!+\!1)(S\!-\!m_2\!+\!1)(S\!-\!m_3)},\\
\beta\!=\!&\sqrt{\!(S\!-\!m_1\!+\!1)(S\!-\!m_2)(S\!-\!m_2\!+\!1)(S\!-\!m_3)},\\
\gamma\!=\!&\sqrt{\!(S\!+\!m_1)(S\!+\!m_2)(S\!+\!m_2\!+\!1)(S\!+\!m_3\!+\!1)},\\
\theta\!=\!&\sqrt{\!(\!S\!-\!m_1\!+\!1\!)\!(\!S\!-\!m_2\!+\!1\!)\!(\!S\!+\!m_2\!+\!1\!)\!(\!S\!+\!m_3\!+\!1\!)\!},
\end{split}
\end{equation}
and
\begin{equation}
\begin{split}
\alpha'\!=\!&\sqrt{\!(S\!-\!m_1\!+\!1)(S\!-\!m_2)(S\!+\!m_2)(S\!+\!m_3\!+\!1)},\\
\beta'\!=\!&\sqrt{\!(S\!+\!m_1)(S\!+\!m_2)(S\!+\!m_2\!+\!1)(S\!+\!m_3\!+\!1)},\\
\gamma'\!=\!&\sqrt{\!(S\!-\!m_1\!+\!1)(S\!-\!m_2)(S\!-\!m_2\!+\!1)(S\!-\!m_3)},\\
\theta'\!=\!&\sqrt{\!(S\!+\!m_1)(S\!-\!m_2)(S\!+\!m_2)(S\!-\!m_3)}.
\end{split}
\end{equation}
When the pair 1,2 is a singlet state, one can find
\begin{equation}
\begin{split}
&\alpha C_{m_1,m_2,m_3}+\beta C_{m_1-1,m_2+1,m_3} \\
=&\gamma C_{m_1,m_2-1,m_3+1}+\theta C_{m_1-1,m_2,m_3+1}\\
=&\alpha' C_{m_1,m_2,m_3}+\beta' C_{m_1-1,m_2+1,m_3}\\
=&\gamma' C_{m_1,m_2-1,m_3+1}+\theta' C_{m_1-1,m_2,m_3+1}=0,
\end{split}
\end{equation}
or when the pair 2,3 is a singlet state, we have
\begin{equation}
\begin{split}
&\alpha C_{m_1,m_2,m_3}+\gamma C_{m_1,m_2-1,m_3+1}\\
=&\beta C_{m_1-1,m_2+1,m_3}+\theta C_{m_1-1,m_2,m_3+1}\\
=&\alpha' C_{m_1,m_2,m_3}+\gamma' C_{m_1,m_2-1,m_3+1}\\
=&\beta' C_{m_1-1,m_2+1,m_3}+\theta' C_{m_1-1,m_2,m_3+1}=0.
\end{split}
\end{equation}
Therefore, $R_i$ and $R'_i$ are positive semidefinite and project out
$\ket{S(i-1,i)}$ and $\ket{S(i,i+1)}$. It may also be possible to rewrite
$R_i$ and $R'_i$ themselves into quadratic forms.

\end{document}


\title{Supplementary Material for\\
``Dimerizations in spin-$S$ antiferromagnetic chains with three-spin
interactions''}
\author{Zheng-Yuan Wang}
\affiliation{Department of Physics, Tokyo Institute of Technology,
Oh-okayama, Meguro-ku, Tokyo, 152-8551, Japan}
\author{Shunsuke C. Furuya}
\affiliation
{DPMC-MaNEP, University of Geneva, 24 Quai Ernest-Ansermet
CH-1211 Geneva, Switzerland}%
\author{Masaaki Nakamura}
\affiliation{
Max-Planck-Institut f\"{u}r
Festk\"{o}rperforschung, Heisenbergstrasse 1, D-70569 Stuttgart, Germany}
\affiliation{Institute of Industrial Science,
the University of Tokyo,
Meguro-ku, Tokyo, 153-8505, Japan}
\author{Ryo Komakura}
\affiliation{Department of Physics, Tokyo Institute of Technology,
Oh-okayama, Meguro-ku, Tokyo, 152-8551, Japan}

\date{\today}
\maketitle

%
In this Supplementary Material, we present detailed calculations to
derive the relations (A3) and (A4) of the main manuscript which are
valid for arbitrary simultaneous eigenstate $\ket{\psi}$ of the
operators $R_2$ and $R'_2$.
%
The operator $R_2$ and $R'_2$ are defined as
\begin{equation}
 R_2 = \frac 12 P_2 +(S+1)Q_2,\quad
  R'_{2} = \frac{1}{2}P_{2}-SQ_{2},
\end{equation}
and the projection operator $P_2$ can be expanded as
\begin{equation}
\begin{split}
\frac{1}{2}P_{2}=&\left[2S(S+1)-1\right]
\left[
2S(S+1)
+2\bm{S}_2\cdot \bm{S}_3
+2\bm{S}_1\cdot \bm{S}_2
\right] \\
&+ S_1^{-} S_3^{+}\left[ S(S+1) - (S_2^{z})^2\right]
+S_1^{+} S_3^{-}\left[ S(S+1) - (S_2^{z})^2\right] \\
& +S_1^{-} S_3^{-} (S_2^{+})^2 + S_1^{+} S_3^{+} (S_2^{-})^2 \\
& +S_1^{-} S_2^{+} S_3^{z}(2S_2^{z}+1)
+S_1^{+} S_2^{-} S_3^{z}(2S_2^{z}-1)\\
& +S_2^{-}S_3^{+}S_1^{z}(2S_2^{z}-1)
+S_2^{+}S_3^{-}S_1^{z}(2S_2^{z}+1) \\
& +4S_1^{z} (S_2^{z})^2 S_3^{z}.
\end{split}
\end{equation}
The terms in $\bra{\psi} R_2 \ket{\psi}$ are calculated in the following
way,
\begin{align}
&\bra{\psi} 2\bm{S}_1 \cdot\bm{S}_2
\ket{\psi} \notag \\ 
=& \sum_{m_1,m_2,m_3}
\left|C_{m_1,m_2,m_3}\sqrt{(S-m_1)(S-m_2+1)}
+
C_{m_1+1,m_2-1,m_3}\sqrt{(S+m_1+1)(S+m_2)} \right|^2 
-2S(S+1) ;\label{eq:S1S2} \\
&\notag \\
&\bra{\psi}S_1^{-} S_3^{+}\left[ S(S+1) - (S_2^{z})^2\right]\ket{\psi}
+
\bra{\psi}S_1^{+} S_3^{-}\left[ S(S+1) - (S_2^{z})^2\right]\ket{\psi}
\notag \\
=&\sum_{m_1',m_2',m_3'} \sum_{m_1,m_2,m_3}
C_{m_1',m_2', m_3'}^* C_{m_1,m_2,m_3}
\bra{m_1',m_2',m_3'}S_1^{-} S_3^{+}\left[ S(S+1) -
 (S_2^{z})^2\right]\ket{m_1,m_2,m_3} 
+\mathrm{H.c.}
\notag \\
=&\sum_{m_1',m_2',m_3'} \sum_{m_1,m_2,m_3}
C_{m_1',m_2', m_3'}^* C_{m_1,m_2,m_3}\delta_{m_1', m_1-1}\delta_{m_3',
 m_3+1}\delta_{m_2', m_2} \notag \\
&\times
\left[ S(S+1) - m_2^2\right]
\sqrt{(S+m_1)(S-m_1+1)}
\sqrt{(S-m_3)(S+m_3+1)}
+\mathrm{H.c.} \notag \\
=&\sum_{m_1,m_2,m_3}
C_{m_1-1,m_2, m_3+1}^* C_{m_1,m_2,m_3}
 \left[ S(S+1) - m_2^2\right]
\sqrt{(S+m_1)(S-m_1+1)}
\sqrt{(S-m_3)(S+m_3+1)}
+\mathrm{H.c.} ; \\
& \notag \\
&\bra{\psi}S_1^{-} S_3^{-} (S_2^{+})^2\ket{\psi}
+
\bra{\psi}S_1^{+} S_3^{+} (S_2^{-})^2\ket{\psi}
\notag \\
=&\sum_{m_1',m_2',m_3'} \sum_{m_1,m_2,m_3}
C_{m_1',m_2', m_3'}^* C_{m_1,m_2,m_3}
\bra{m_1',m_2',m_3'}
S_1^{-} S_3^{-} (S_2^{+})^2
\ket{m_1,m_2,m_3}
+\mathrm{H.c.}
\notag \\
=&\sum_{m_1',m_2',m_3'} \sum_{m_1,m_2,m_3}
C_{m_1',m_2', m_3'}^* C_{m_1,m_2,m_3}
\delta_{m_1', m_1-1}
\delta_{m_3', m_3-1}
\delta_{m_2', m_2+2}\notag \\
&\times
\sqrt{(S+m_1)(S-m_1+1)}
\sqrt{(S+m_3)(S-m_3+1)} 
\sqrt{(S-m_2-1)(S-m_2)(S+m_2+1)(S+m_2+2)} \notag \\
&+\mathrm{H.c.} \notag
\end{align}
\begin{align}
=&\sum_{m_1,m_2,m_3} 
C_{m_1-1,m_2+2, m_3-1}^* C_{m_1,m_2,m_3} \notag \\
& \times
\sqrt{(S+m_1)(S-m_1+1)}
\sqrt{(S+m_3)(S-m_3+1)}
\sqrt{(S-m_2-1)(S-m_2)(S+m_2+1)(S+m_2+2)}
\notag \\
&+\mathrm{H.c.} ; \\
&\phantom{\sum_{m_1,m_2,m_3}
C_{m_1-1,m_2, m_3+1}^* C_{m_1,m_2,m_3}
 \left[ S(S+1) - m_2^2\right]
\sqrt{(S+m_1)(S-m_1+1)}
\sqrt{(S-m_3)(S+m_3+1)}
+\mathrm{H.c.} ;}\notag \\
&\bra{\psi}S_1^{-} S_2^{+} S_3^{z}(2S_2^{z}+1)\ket{\psi}
+
\bra{\psi}S_1^{+} S_2^{-} S_3^{z}(2S_2^{z}-1)\ket{\psi}
\notag \\
=&\sum_{m_1',m_2',m_3'} \sum_{m_1,m_2,m_3}
C_{m_1',m_2', m_3'}^* C_{m_1,m_2,m_3}
\bra{m_1',m_2',m_3'}S_1^{-} S_2^{+} S_3^{z}(2S_2^{z}+1)
\ket{m_1,m_2,m_3}+\mathrm{H.c.}
\notag \\
=&\sum_{m_1',m_2',m_3'} \sum_{m_1,m_2,m_3}
C_{m_1',m_2', m_3'}^* C_{m_1,m_2,m_3} \delta_{m_1', m_1-1}
\delta_{m_3', m_3}
\delta_{m_2', m_2+1}\notag \\
&\times
\sqrt{(S+m_1)(S-m_1+1)}
\sqrt{(S-m_2)(S+m_2+1)}
 m_3 (2m_2+1) + \mathrm{H.c.}
\notag \\
=&\sum_{m_1,m_2,m_3}
C_{m_1-1,m_2+1, m_3}^* C_{m_1,m_2,m_3}
\sqrt{(S+m_1)(S-m_1+1)}
\sqrt{(S-m_2)(S+m_2+1)}
 m_3 (2m_2+1)
+\mathrm{H.c.};  \\
& \notag \\
&\bra{\psi}S_2^{-}S_3^{+}S_1^{z}(2S_2^{z}-1)\ket{\psi}
+
\bra{\psi}S_2^{+}S_3^{-}S_1^{z}(2S_2^{z}+1)\ket{\psi}
\notag \\
=&\sum_{m_1',m_2',m_3'} \sum_{m_1,m_2,m_3}
C_{m_1',m_2', m_3'}^* C_{m_1,m_2,m_3}
\bra{m_1',m_2',m_3'}
S_2^{-}S_3^{+}S_1^{z}(2S_2^{z}-1)
\ket{m_1,m_2,m_3}+\mathrm{H.c.}
\notag \\
=&\sum_{m_1',m_2',m_3'} \sum_{m_1,m_2,m_3}
C_{m_1',m_2', m_3'}^* C_{m_1,m_2,m_3}\delta_{m_2', m_2-1}
\delta_{m_3', m_3+1}
\delta_{m_1', m_1}
\notag \\
&\times
\sqrt{(S+m_2)(S-m_2+1)}
\sqrt{(S-m_3)(S+m_3+1)}
 m_1 (2m_2-1) + \mathrm{H.c.}
\notag \\
=&\sum_{m_1,m_2,m_3}
C_{m_1,m_2-1, m_3+1}^* C_{m_1,m_2,m_3}
\sqrt{(S+m_2)(S-m_2+1)}
\sqrt{(S-m_3)(S+m_3+1)}
 m_1 (2m_2-1)
+\mathrm{H.c.} ; \\
\notag \\
&\bra{\psi}
4S_1^{z} (S_2^{z})^2 S_3^{z}
\ket{\psi}\notag \\
=&\sum_{m_1,m_2,m_3}
4|C_{m_1,m_2, m_3}|^2m_1(m_2)^2m_3.
\end{align}
Calculation of $\bra{\psi}Q_2\ket{\psi}$ is performed in the same manner
as for Eq.~\eqref{eq:S1S2}.

After all, one finds the following relation.
\begin{align}
\braket{\psi|R_2|\psi}=
&S(2S+3)
 \biggl\{ 2S(S+1)+2\sum_{m_1,m_2,m_3} |C_{m_1,m_2,m_3}|^2 m_1m_2 \notag
 \\ 
& \quad +\sum_{m_1,m_2,m_3}
(C_{m_1-1,m_2+1,m_3}^*C_{m_1,m_2,m_3}+C_{m_1-1,m_2+1,m_3}C_{m_1,m_2,m_3}^*)
 \notag \\
& \qquad \times\sqrt{(S+m_1)(S-m_1+1)}\sqrt{(S-m_2)(S+m_2+1)}\notag \\
& \quad +2 \sum_{m_1,m_2,m_3} 
|C_{m_1,m_2,m_3}|^2 m_2m_3 \notag \\
& \quad + \sum_{m_1,m_2,m_3}
(C_{m_1,m_2-1,m_3+1}^*C_{m_1,m_2,m_3}+C_{m_1,m_2-1,m_3+1}C_{m_1,m_2,m_3}^*)\notag \\
& \qquad \times\sqrt{(S+m_2)(S-m_2+1)}\sqrt{(S+m_3+1)(S-m_3)} \biggr\}
 \notag \\
&\phantom{\sum_{m_1,m_2,m_3}
\left|
\alpha C_{m_1,m_2,m_3}
+\beta C_{m_1-1,m_2+1,m_3}
+\gamma C_{m_1,m_2-1,m_3+1}
+\theta C_{m_1-1,m_2,m_3+1}
\right|^2,}\notag
\end{align}
\begin{align}
 \phantom{\braket{\psi|R_2|\psi}=}
&+\sum_{m_1,m_2,m_3}
(C_{m_1-1,m_2, m_3+1}^* C_{m_1,m_2,m_3} +C_{m_1-1,m_2, m_3+1}
 C_{m_1,m_2,m_3}^* ) \notag \\
& \quad \times \left[ (S+1)^2 - m_2^2\right]
\sqrt{(S+m_1) (S+m_3+1)}
\sqrt{(S-m_3) (S-m_1+1)} \notag \\
&+\sum_{m_1,m_2,m_3}
(C_{m_1-1,m_2+1, m_3}^* C_{m_1,m_2-1,m_3+1}+C_{m_1-1,m_2+1,
 m_3}C_{m_1,m_2-1,m_3+1}^*) \notag \\
& \quad \times\sqrt{(S+m_1) (S+m_2) (S-m_2+1) (S+m_3+1)} \notag \\
& \quad \times\sqrt{ (S-m_1+1) (S-m_2)(S+m_2+1) (S-m_3)} \notag \\
&+\sum_{m_1,m_2,m_3}
(C_{m_1-1,m_2+1, m_3}^* C_{m_1,m_2,m_3}+C_{m_1-1,m_2+1, m_3}
 C_{m_1,m_2,m_3}^*) \notag \\
& \quad \times
\sqrt{(S+m_1)(S+m_2+1)}
\sqrt{(S-m_2)(S-m_1+1)}
 m_3 (2m_2+1)
\notag \\
&+\sum_{m_1,m_2,m_3}
(C_{m_1,m_2-1, m_3+1}^* C_{m_1,m_2,m_3}+C_{m_1,m_2-1, m_3+1}^*
 C_{m_1,m_2,m_3})\notag \\
& \quad \times
\sqrt{(S+m_2)(S+m_3+1)}
\sqrt{(S-m_3)(S-m_2+1)}
 m_1 (2m_2-1)
\notag \\
&+\sum_{m_1,m_2,m_3}4C_{m_1,m_2, m_3}^* C_{m_1,m_2,m_3}m_1(m_2)^2m_3
 \notag \\
=&\sum_{m_1,m_2,m_3}
\left|
\alpha C_{m_1,m_2,m_3}
+\beta C_{m_1-1,m_2+1,m_3}
+\gamma C_{m_1,m_2-1,m_3+1}
+\theta C_{m_1-1,m_2,m_3+1}
\right|^2,
\end{align}
where $\alpha$, $\beta$, $\gamma$ and $\theta$ are defined as
\begin{equation}
\begin{split}
&\alpha=\sqrt{(S+m_1)(S+m_2+1)(S-m_2+1)(S-m_3)}, \\
&\beta=\sqrt{(S-m_1+1)(S-m_2)(S-m_2+1)(S-m_3)}, \\
&\gamma=\sqrt{(S+m_1)(S+m_2)(S+m_2+1)(S+m_3+1)}, \\
&\theta=\sqrt{(S-m_1+1)(S-m_2+1)(S+m_2+1)(S+m_3+1)}.
\end{split}
\end{equation}
In a similar way,
$\bra{\psi}R'_{2}\ket{\psi}$ can be calculated as
\begin{align}
\braket{\psi|R'_{2}|\psi}
=
&\left( 2S^2+S-1\right)
 \biggl\{ 2S(S+1)+2\sum_{m_1,m_2,m_3} |C_{m_1,m_2,m_3}|^2 m_1m_2 \notag
 \\ 
&\quad +\sum_{m_1,m_2,m_3}
(C_{m_1-1,m_2+1,m_3}^*C_{m_1,m_2,m_3}
 +C_{m_1-1,m_2+1,m_3}C_{m_1,m_2,m_3}^*)  \notag \\
&\qquad \times\sqrt{(S+m_1)(S-m_1+1)}\sqrt{(S-m_2)(S+m_2+1)}\notag \\
&\quad +2 \sum_{m_1,m_2,m_3} |C_{m_1,m_2,m_3}|^2 m_2m_3 \notag \\
&\quad + \sum_{m_1,m_2,m_3}
(C_{m_1,m_2-1,m_3+1}^*C_{m_1,m_2,m_3}+C_{m_1,m_2-1,m_3+1}C_{m_1,m_2,m_3}^*)\notag \\
&\qquad \times\sqrt{(S+m_2)(S-m_2+1)}\sqrt{(S+m_3+1)(S-m_3)}
 \biggr\}\notag \\
&+\sum_{m_1,m_2,m_3}
(C_{m_1-1,m_2, m_3+1}^* C_{m_1,m_2,m_3} +C_{m_1-1,m_2, m_3+1}
 C_{m_1,m_2,m_3}^* )\notag \\
&\quad \times \left( S^2 - m_2^2\right)
\sqrt{(S+m_1) (S+m_3+1)}
\sqrt{(S-m_3) (S-m_1+1)}\notag \\
&+\sum_{m_1,m_2,m_3}
(C_{m_1-1,m_2+1, m_3}^* C_{m_1,m_2-1,m_3+1}+C_{m_1-1,m_2+1,
 m_3}C_{m_1,m_2-1,m_3+1}^*)\notag \\
&\quad \times\sqrt{(S+m_1) (S+m_2) (S-m_2+1) (S+m_3+1)}\notag \\
&\quad \times\sqrt{ (S-m_1+1) (S-m_2)(S+m_2+1) (S-m_3)}\notag \\
&\phantom{\sum_{m_1,m_2,m_3}
\left|
\alpha' C_{m_1,m_2,m_3}
+\beta' C_{m_1-1,m_2+1,m_3}
+\gamma' C_{m_1,m_2-1,m_3+1}
+\theta' C_{m_1-1,m_2,m_3+1}
\right|^2,}\notag
\end{align}
\begin{align}
\phantom{\braket{\psi|R'_{2}|\psi}
=}
&+\sum_{m_1,m_2,m_3}
(C_{m_1-1,m_2+1, m_3}^* C_{m_1,m_2,m_3}+C_{m_1-1,m_2+1, m_3}
 C_{m_1,m_2,m_3}^*)\notag \\
&\quad \times
 \sqrt{(S+m_1)(S+m_2+1)}
\sqrt{(S-m_2)(S-m_1+1)}
 m_3 (2m_2+1) \notag \\
&+\sum_{m_1,m_2,m_3}
(C_{m_1,m_2-1, m_3+1}^* C_{m_1,m_2,m_3}+C_{m_1,m_2-1, m_3+1}^*
 C_{m_1,m_2,m_3})\notag \\
&\quad \times
\sqrt{(S+m_2)(S+m_3+1)}
\sqrt{(S-m_3)(S-m_2+1)}
 m_1 (2m_2-1) \notag \\
&+\sum_{m_1,m_2,m_3}4C_{m_1,m_2, m_3}^*
 C_{m_1,m_2,m_3}m_1(m_2)^2m_3\notag \\
=&\sum_{m_1,m_2,m_3}
\left|
\alpha' C_{m_1,m_2,m_3}
+\beta' C_{m_1-1,m_2+1,m_3}
+\gamma' C_{m_1,m_2-1,m_3+1}
+\theta' C_{m_1-1,m_2,m_3+1}
\right|^2,
\end{align}
where
\begin{equation}
\begin{split}
&\alpha'=\sqrt{(S-m_1+1)(S-m_2)(S+m_2)(S+m_3+1)}, \\
&\beta'=\sqrt{(S+m_1)(S+m_2)(S+m_2+1)(S+m_3+1)}, \\
&\gamma'=\sqrt{(S-m_1+1)(S-m_2)(S-m_2+1)(S-m_3)}, \\
&\theta'=\sqrt{(S+m_1)(S-m_2)(S+m_2)(S-m_3)}.
\end{split}
\end{equation}

%